\documentstyle[preprint,aps,amssymb,prb,epsf]{revtex}
\tightenlines
\newcommand{\beq}{\begin{equation}}
\newcommand{\eeq}{\end{equation}}
\newcommand{\bea}{\begin{eqnarray}}
\newcommand{\eea}{\end{eqnarray}}
\begin{document}

\title{{\it  Article submitted for the special issue of the American Journal
of Physics concerning Thermodynamics, Statistical Mechanics, and Statistical
Physics, to be edited by Jan Tobochnik and Harvey Gould}\\
Atoms in nanotubes: small dimensions and variable
dimensionality}

\author{George Stan,$^1$ Silvina M. Gatica,$^{1,2}$ Massimo
Boninsegni,$^3$\\ Stefano Curtarolo,$^{1,4}$ and Milton W. Cole$^{1,5}$}

\address{$^1$Department of Physics and Center for Materials Physics,\\
Pennsylvania State University, University Park, PA 16802\\
$^2$Departamento de F$\acute{\imath}$sica,
Universidad de Buenos Aires, Buenos Aires 1428, Argentina\\
$^3$Department of Physics, San Diego State University, San Diego,
CA 92182\\
$^4$Present address: Deptartment of Materials Science and Engineering, MIT,
Cambridge,\\ MA 02139\\
$^5${corresponding author, e-mail: mwc@psu.edu}}

\maketitle

\medskip \noindent Newly discovered carbon nanotubes provide an
environment in which small atoms move relatively freely. An
assembly of such atoms provides a realization of a quasi-one
dimensional system which is an ideal testing ground for concepts
and mathematics of statistical physics.

\medskip \noindent {\bf I. INTRODUCTION}

One of the recurrent themes in statistical physics is the
sensitivity of natural phenomena to the spatial dimensionality of
the system. The properties of
three-dimensional (3D) matter (for example, solid argon)
differ quantitatively from 2D matter made of
the same constituent particles (for example, a layer of argon, one
atom thick, deposited on a graphite surface). What may not be
obvious is that the behavior of 3D systems is {\it qualitatively}
different from that of 2D systems. For
example, a monolayer film has properties which are sensitive to the
underlying solid, such as its lattice constant. This film may
exhibit ordered phases which have no 3D counterpart, such as
commensurate solids.\cite{bruch} The collective modes of the
system (for example, phonons) also will produce thermodynamic
properties which depend on the dimensionality. More subtle
are differences in the values of critical exponents which
characterize thermodynamic singularities near phase
transitions. These differences reflect the fact that the
correlations responsible for order are more sensitive to
fluctuations in 2D than in 3D. This sensitivity is even more
dramatic in 1D systems,\cite{gubbins} and one consequence of these
fluctuations is that no ordered phases may exist in 1D at finite
temperature
$T$.\cite{landau}

One might wonder whether 1D systems exist in nature which can test
this intriguing ``null'' hypothesis. This search has led us
to explore the statistical mechanics of atoms moving
in carbon nanotubes (see
Fig.~\ref{fig:nanotubes}).\cite{stan1,stan2,vidales,stan3} We have
found a remarkable variety of phenomena which exhibit a range of
effective dimensionalities, depending on the thermodynamic
variables (number of atoms $N$ and temperature $T$), microscopic
variables (atomic size relative to nanotube radius), and the
geometry (isolated tubes or ordered arrays of tubes). This paper
discusses some aspects of the behavior that we have found,
revealing the system to be a marvelous playground for concepts of
thermal physics.

The field of carbon nanotubes was born in 1991 with the discovery
by Iijima\cite{iijima} and the determination of conditions
for the synthesis of large quantities of nanotubes.\cite{ebbesen}
The tubes consist of graphite sheets rolled up into hollow
cylinders, and they appear as either single or coaxial cylinders,
called single-wall and multi-wall nanotubes, respectively
(see Fig.~\ref{fig:nanotubes}). A remarkable feature is their
width-to-length ratio, due to their very small diameters ($\sim 14\,
\AA$) and relatively long lengths (on the order of tens of
$\mu$m).\cite{white} Recent experiments have successfully produced
bundles of single-wall nanotubes,\cite{smalley} arranged in
triangular lattices with a lattice constant of $17\, \AA$. The
distribution of radii of the tubes is narrowly peaked around $7\,
\AA$. This size is such that many species of small atoms (diameter of order 
$3 \AA$) fit comfortably within the tube and are therefore strongly imbibed 
by these tubes.\cite{H_adsorption} In the nanotube bundle geometry we
expect that small atoms are even more strongly adsorbed in the
narrow interstitial channels between the tubes than within
them.\cite{stan3,mays} Indeed, a recent measurement for $^4$He found an
interstitial binding energy which is 2.3 times as large as the binding
energy on the basal plane of graphite (which is the highest for any planar
surface).\cite{teizer}

The outline of the paper is the following. Sections~II and III
discuss the properties of weakly interacting quantum gases confined
within these tubes. We find that the thermal behavior as a function
of T and N corresponds to dimensions $D = 1, 2$, or 3. Section~IV discusses 
the results of a numerical study of a strongly interacting fluid (helium), 
which exhibits extraordinary quasi-1D behavior (with a transition) when 
confined interstitially within a bundle of nanotubes. Section~V discusses a
dimensional crossover manifested by the phonons of a high density
film adsorbed within a tube. Section~VI summarizes  these
phenomena. This variety of behavior provides intriguing
manifestations of novel physics and stimulating examples of the
beauty present in the diverse field of statistical physics.

\medskip \noindent {\bf II. ULTRALOW DENSITY GAS}

One of the simplest models in statistical physics is the
classical ideal gas. May such a simple model be
applied to atoms moving within the extremely anisotropic,
confined geometry of a nanotube? The answer is yes, and the domain
of classical ideal gas behavior is quite extended. The problem is
interesting because it involves the quantum-mechanical spectrum of
the individual particles, even though classical statistics describes
the thermal properties of the gas as a whole. A
surprising fact is that at any nonzero temperature
there exists a density regime in which the classical ideal
gas approximation is valid.

The classical ideal gas assumption implies that the probability
that a single particle has energy $E$ is proportional to
the Boltzmann function,
$p(E) = e^{-\beta E}$, where $\beta^{-1} = k_B T$. The
mean energy
$U$ of the system of $N$ atoms can be found from $n(E)$, the density
of states, and the average energy per particle
$\langle E \rangle=U/N$ is given by
\beq
\langle E \rangle =
\frac{\int \! dE\, E\, n(E)\, p(E)}{\int \! dE\, n(E)\,
p(E)} .
\eeq
The behavior of this system resides in the function
$n(E)$, which we now consider for gases inside nanotubes.

An
adequate approximation of the energy spectrum can be found by
assuming that the atom's potential energy is a function
only of the radial coordinate
$r$. Then the Schr\"{o}dinger equation is separable, and the
resulting wave functions may be written as
\beq
\Psi({\bf r}) = L^{-1/2}\, e^{ikz}\, \psi_{n \nu}(r,\varphi) ,
\eeq
where $E =\hbar^2 k^2/2m + \epsilon_{n \nu}$, $\epsilon_{n
\nu}$ is the energy of transverse motion, $L$ is the tube length,
$\varphi$ is the azimuthal angle, and $k$ is the axial wave vector.
Figure~\ref{fig:pot} displays a model calculation of the radial
dependence of the potential energy of a He atom inside a nanotube
of radius $5\, \AA$. Also shown are the lowest lying eigenfunctions
for motion in the plane perpendicular to the axis (the
determination of the eigenfunctions is discussed in the Appendix ).
These eigenfunctions involve azimuthal excitation and are labelled
by a quantum number $\nu$, corresponding to wave functions
\beq
\psi_{1 \nu} (r,\varphi) = f_{1 \nu} (r) e^{i \nu \varphi} ,
\eeq
where $\nu$ is an integer, and $f_{1 \nu}(r)$ is the radial wave
function. Because of the small tube size, the radial degree of
freedom has a large excitation energy. Hence, at low temperature we
may neglect the thermal contribution from all radially excited
states and consider only  states with $n=1$. Under these
circumstances, the effective dimensionality is two, arising from
the azimuthal and longitudinal excitations, respectively.
Intriguing behavior arises when we address the axial motion.
The density
of states for 1D motion alone is\cite{dos}
\bea
n_{\rm axial}(E) = b\, E^{-1/2}, \qquad
b = (g L / \hbar)\, (2m)^{1/2} ,
\label{eq:1d_dos}
\eea
where $g$ is the spin degeneracy. Together with the azimuthal
excitation, we now have two ``active'' degrees of freedom. The
total density of states is
\beq
n(E) = \sum_\nu b\, (E - \epsilon_{1 \nu})^{-1/2}\,H(E -
\epsilon_{1 \nu}) ,
\label{eq:dens_states}
\eeq
where $H$ is the Heaviside step function and the sum is over
$\nu = 0, \pm 1, \pm 2, \ldots$ The states with $\nu \neq 0$ are
doubly degenerate, while the $\nu=0$ state is nondegenerate. 
Figure~\ref{fig:dens_states} shows $n(E)$, which has a serrated
shape due to the onset of successive azimuthal contributions. At
very low $T$, only the 1D motion, corresponding to $\nu = 0$, is
excited, while higher $\nu$ values become relevant at higher $T$.
Figure~\ref{fig:heat_c} shows the dimensionless specific heat $c
\equiv C/N k_B$ calculated from this spectrum for a
system of $N$ particles. Note that the low $T$ behavior exhibits a
specific heat equal to the 1D classical ideal gas value ($c=1/2$),
while the high
$T$ limit is the 2D value ($c=1$). The overshoot behavior in the
intermediate crossover region of $T$ in Fig.~\ref{fig:heat_c}
($c>1$) is reminiscent of that seen in the specific heat of
diatomic molecules, \cite{note2} which is expected because of a
qualitatively similar excitation spectrum.\cite{note3} The
dimensionality crossover occurs at  a value of $T$
corresponding approximately to the first azimuthal excitation
energy. As seen in Fig.~\ref{fig:heat_c}, this value of $T$ depends
on the nanotube radius $R$, as a greater $R$ implies smaller
azimuthal energy. Figure~\ref{fig:dimens} shows how the effective
dimensionality crossover depends on $R$ and $T$ for helium. The
effective dimensionality variation (from one to three as $T$ is
increased) is a general feature of the system.

We now address the issue of the validity of the classical ideal gas
approximation. At sufficiently high $T$, quantum corrections can be
evaluated from a virial (Wigner-Kirkwood) expansion, as is familiar
from statistical physics. Quantum statistical
corrections become important when the de
Broglie thermal wave length $\lambda_T= (2 \pi \beta
\hbar^2/m)^{1/2}$ becomes comparable to the interparticle spacing
($L/N$) along the tube. More interesting perhaps is the
effects of interactions. There are two heuristic arguments which
imply that (even at very low $T$) there will always exist a
density below which the classical ideal limiting behavior occurs. 
One is simply that the quasi-1D system does not condense (at finite
$T$). This behavior is nearly unique; to our
knowledge, analogous (equilibrium) behavior for neutral particles
exists only for spin polarized atomic hydrogen.\cite{nosanow} The
second argument pertains to the effect of the ``hard-core''
repulsion between atoms. We will show elsewhere\cite{boninsegni} that the 
spread of the single particle wave functions around the perimeter of the 
tube is sufficient to reduce the core effect, that is, allow atoms to pass by 
one another.

\medskip \noindent {\bf III. FINITE DENSITY IDEAL QUANTUM GAS}

At a finite density we need to evaluate the effect of quantum
statistics even if the gas is noninteracting, because at low $T$
the system becomes degenerate. The
calculational procedure is a straightforward extension of that
conventionally used for 3D translationally invariant systems. We
must determine the energy of the system
\beq
U = \! \int \! dE \, E\, n(E)\, f_{\pm}(E)
\eeq
where $f_{\pm}(E)$ is the Bose or Fermi occupation function:
\beq
f_{\pm}(E) = {1 \over e^{\beta (E-\mu)} \pm 1} .
\eeq
The chemical potential $\mu$ is implicitly determined by the
number of  atoms:
\beq
N = \! \int \! d E\, n(E)\, f_{\pm}E) .
\eeq
We have performed this numerical procedure for Fermi systems having
various numbers of particles $N$. Each value of $N$ yields a
different Fermi energy determined by
\beq
N= \! \int^{E_F} \! d E\, n(E) .
\eeq
For convenience, we shift the zero of energy to be
$\epsilon_{1 0}$, the lowest azimuthal energy. For $R =
5\, \AA$ and
$^3$He, the first excited state, $\epsilon_{1 1}$, is at an energy
$\Delta/k_B = 2.3\, {\rm K}$, as seen in Fig.~\ref{fig:dens_states}.
The three examples we will discuss have Fermi energies
$E_F = 0.05$, 0.95, and $1.05\, \Delta$, respectively, as indicated
in Fig.~\ref{fig:dens_states}. These values were chosen to
demonstrate the variety of crossover (classical/quantum and 1D/2D)
behavior in this system.

At each density, the system remains classical down to a degeneracy
temperature $T_d = E_F/k_B$. This behavior is most clearly seen in
Fig.~\ref{fig:deg_Fermi} for the low density ($E_F=0.05\,\Delta$)
case, for which $T_d \sim 0.1\, {\rm K}$. For the two higher
densities, the higher degeneracy temperature implies that the
classical regime appears only at higher $T$, that is, above $T \sim
2\, {\rm K}$, as seen in Fig.~\ref{fig:deg_Fermi}. Above that
value, the effect of the statistics  is negligible, that is, the
results of Section~II apply (as in Fig.~\ref{fig:heat_c}). As $T$
falls below
$T_d$, quantum effects drive the specific heat to zero; the
extreme degenerate behavior of a quantum system is given in terms
of the density of states at the Fermi energy:\cite{lowT_cv}
\beq
C_V/k_B \simeq \frac{\pi^2}{3}\,n(E_F)\,k_B \label{eq:heat_caplowT} 
T .
\eeq

In the degenerate regime of very low $T$,
Eq.~(\ref{eq:heat_caplowT}) applied to our three densities yields
dramatically different slopes at low $T$, as is consistent with the
infinite discontinuity in the density of states at the threshold
for exciting the first azimuthal level (see
Eq.~(\ref{eq:dens_states}) and Fig.~\ref{fig:dens_states}).
Specifically, at $E_F=1.05\,\Delta$ the low $T$ specific heat is a
factor of ten higher than at $E_F=0.95\,\Delta$, while it is
approximately double that at
$E_F=0.05\,\Delta$.

Note the curves in Fig. 6 at finite density have values of $C/N k_B$ which are
below the classical curve of Fig. 4. This behavior is a consequence of the 
fact that the quantum energy must coincide with the classical energy at 
high $T$. Hence the integrated heat capacity difference satisfies
\beq
\int_0^\infty \! d T\, (C - C_{\rm classical}) = U_0,
\eeq
where $U_0$ is the energy of the system at $T = 0$, corresponding to
all states below the Fermi energy being fully occupied. At high $N$,
the right side is very large, so the ``quantum deficiency'' 
$C- C_{\rm classical}$ must be very large.

Qualitatively similar effects of quantum degeneracy occur for Bose
systems. At high $T$, the system deviates from the classical ideal
gas when $L/N$ becomes comparable to the de Broglie wave length,
that is, when the delocalized single-particle wave functions
overlap. We focus, instead, on the very low $T$ regime. Here one
finds the following behavior by applying conventional low $T$
expansion techniques. The low-$T$ specific heat can be shown to
satisfy
\beq
C/N k_B = (3 \zeta (3/2)/4) g L /N \lambda_T , \label{eq:zeta}
\eeq
where $\zeta(x)$ is the Riemann zeta function.
Equation~(\ref{eq:zeta}) holds in the regime where the right-hand
side is much less than one. Therefore, in this regime, the
specific heat scales as T$^{1/2}$. It is amusing to find the
general behavior of the specific heat of a Bose gas to be of the
form
$T^{d/2}$, even in the anomalous ($d > 2$) case of Bose-Einstein
condensation. The expression for each dimensionality has
essentially the same form, with the rightmost term raised to the
power $d$.\cite{gunton}

At low $T$ the interactions between real particles usually cannot be
neglected and the ideality assumption fails (except at very low
density). Phonon-like collective modes appear in the Fermi and the
Bose cases,\cite{lieb} while spin waves appear in the case of spin,
corresponding to the realization of a so-called Luttinger
liquid.\cite{luttinger} To our knowledge, this concept has not yet
been applied to the nanotube problem.

\medskip \noindent {\bf IV. INTERACTING FLUIDS}

We now consider systems of interacting atoms confined to within
nanotubes. In some cases, these may be adequately described by
classical statistical mechanics, while in others cases a quantum
treatment is necessary. The division between these cases is
conventionally made by evaluating the De Boer quantum parameter
$\Lambda^*$, which is a dimensionless measure of the ratio of
quantum kinetic energy to potential energy: \cite{quantum_KE} 
\beq
\Lambda^* = h/\sigma(m \varepsilon)^{1/2} .
\eeq
Here $\sigma$ and $\varepsilon$ are parameters of the Lennard-Jones
interaction between atoms of mass $m$:
\beq
V = 4 \varepsilon [(\sigma/r)^{12} - (\sigma/r)^6] .
\eeq
When $\Lambda^*$ is ``small,'' classical theory is usually
applicable, while otherwise quantum effects are not negligible. For
example, argon ($\Lambda^* = 0.063$) and heavier inert gases are
classical in their behavior. The light and weakly interacting
systems H$_2$ ($\Lambda^* = 1.72$) and $^4$He ($\Lambda^* = 3.2$)
are fully quantum mechanical, while neon ($\Lambda^* = 0.58$) is a
borderline case of modest quantum corrections to the thermal
properties.\cite{wang}

Little theoretical work has been carried out so far for interacting
atoms in single nanotubes. There is,
however, a related situation for which some interesting results
exist. That is the case of an ordered array of nanotubes, which
form a close-packed bundle similar to strands of
bucatini\cite{note4} held tightly in one's hand. We have found that
tiny atoms are strongly attracted to the interstices between the
tubes; these spaces are sufficiently small that a purely 1D theory
may be applicable.\cite{stan3} This limit conveniently permits
clean statistical mechanical calculations. For example, there is
the venerable problem of
the 1D classical gas of particles of specified hard core diameter
$a$, which was solved by Tonks and Langmuir.\cite{tonks} The 1D
pressure satisfies
\beq
p = k_B T/(L/N - a) .
\eeq
Interestingly, the classical system lends itself to an exact
analytical treatment even in the case of more general interactions,
as long as only nearest neighbors interact.\cite{n.n.interactions}

From a fundamental point of view, the most interesting
case may be the extreme quantum case of He.
Figure~\ref{fig:bound_state} shows the energy per atom of $^4$He in
1D at $T=0$, computed with the Diffusion Monte Carlo (DMC)
method.\cite{schmidt} The data in the inset indicate that there may exist a 
low density bound state; that is, the 1D ground state is a liquid. This finding
 has been corroborated by a recent variational calculation of Krotscheck and 
Miller.\cite{krotscheck} The binding energy is very small, of the order of 
$10\, {\rm mK}$ per atom, at a mean interatomic spacing of about $15 \AA$. 
Such a small cohesive energy is even less than what one might have expected 
from an extrapolation of the decrease in binding known to occur on reducing
from 3D ($E/N = -7.2 K$) to 2D ($E/N = -0.8 K$) liquid
He.\cite{note5,whitlock}

As indicated in Section~I, there can be no phase transition in
1D at finite $T$; this implies that the condensed state is absent
except at $T=0$. A fragile liquid can exist, however, at $T \neq
0$, in a nanotube bundle array, due to the cooperative attractive
interactions from atoms in neighboring tubes. Such a novel
anisotropic quantum fluid would be of immense fundamental interest.
If one recalls\cite{kosterlitz} that 2D Bose systems exhibit
superfluidity without Bose condensation, one wonders what might be
the properties of this quasi-1D fluid. No research has been done
thus far to explore this question. 

\medskip \noindent {\bf V. DIMENSIONALITY CROSSOVER OF THE PHONONS}

The previous discussion pertains to the regime when the adsorbed
material is a fluid. At high density, there are other interesting
possibilities. One example involves a bundle of nanotubes (because
it provides a 3D environment), which can permit the formation of
solids within the tubes. Atoms of diameters $\sim 3 \AA$ will form
a cylindrical shell film coating the nanotube wall. If we turn to
the thermal excitation of this solid, we encounter another instance
of varying effective dimensionality. This is the case of phonon
excitation. We have studied\cite{vidales} the specific heat arising
from the phonons (assuming that the film's atoms lie on a cylinder
of radius
$R$ and possess an isotropic speed of sound $s$). The calculation
evaluates the total energy
\beq
U = \sum_\alpha \frac{\hbar \omega_\alpha}{e^{\beta \hbar
\omega_\alpha}- 1} ,
\eeq
where $\omega_\alpha$ is the frequency of phonon mode $\alpha$ (which, in
general, is characterized by axial and azimuthal contributions). The results
are shown in Figs.~\ref{fig:phonons_crossover} and
\ref{fig:phonons_heat_c}. Figure~\ref{fig:phonons_crossover} shows
the relationship between a dimensionless function proportional to
$C/T$
\beq
F(T)=( C/k_B) (\lambda/L) ,
\label{eq:heat_cap} \eeq
and a dimensionless temperature
\beq
t=R/\lambda
\label{eq:t} \eeq
Here
\beq
\lambda=\hbar \beta s
\eeq
is the thermal phonon wave length at temperature $T$. Note that the
heat capacity is linear in $T$ at low temperature while the high
temperature result is quadratic in $T$. This crossover is analogous
to that found for the classical ideal gas: at low $T$ only phonons
having wave vectors directed along the axis are excited, while at
high $T$ the azimuthal modes also become excited. The effective
dimensionality is seen in Fig.~\ref{fig:phonons_crossover} to
change from one to two when R becomes about $0.15 \lambda$. This
value is where the circumference becomes nearly equal to a thermal
wave length, which is a very logical onset condition for the
azimuthal degree of freedom. Fig.~\ref{fig:phonons_heat_c} shows
how the 1D and 2D regimes depend on the specific material; this
behavior depends only on the value of $s$, which is a function of
film density.

\medskip \noindent {\bf VI. CONCLUSIONS}

From the preceding examples it is evident that atoms in nanotubes
exhibit unusual properties which fascinate and stimulate our
imagination. We have only briefly alluded to a subset of the wide
variety of simple research problems which are yet to be thoroughly
explored. Although many of these problems involve numerical methods,
others are amenable to simple modelling. These should provide an
inspiration to both students of statistical physics and researchers
in condensed matter physics.

\medskip \noindent {\bf ACKNOWLEDGEMENTS}

We would like to thank Aldo Migone, Moses Chan, Jainendra Jain,
Yong-Baek Kim, Karl Johnson, A. T. Johnson, Ana Maria Vidales, and
Vincent Crespi for very useful discussions. We are happy to
acknowledge support from the National Science Foundation under
research grants DMR-9705270 and DMR-9802803 and ACS-PRF
\#31641-AC5 from the Petroleum Research Fund of the American
Chemical Society.

\medskip \noindent {\bf APPENDIX. DETERMINATION OF THE WAVE
FUNCTIONS}

For a potential possessing cylindrical symmetry, the radial 
Schr\"odinger equation becomes
\begin{equation}
\frac{d^2 f}{dr^2} + \frac{1}{r} \frac{df}{dr} +
\biggl\{\frac{2m}{\hbar^2}\left[\epsilon _{n\nu} - V(r)\right] -
\frac{\nu^2}{r^2} \biggr\} f = 0 ,
\label{eq:rad_sch}
\end{equation}
where $f(r)$ is the radial wave function. In general, solving
Eq. \ref{eq:rad_sch} requires using a numerical method, for example the Numerov
method. Instead, a simple way of estimating the energy levels 
$\epsilon_{n\nu}$ is to use a piecewise constant model
potential\cite{stan1} constructed as shown in Fig.~\ref{fig:pot}.
Let $V(r)=V_0$ for
$0<r<a$, where $a$ is the inner distance at which the potential is
$(V_{\rm min}+V_0)/2$;
$V_{\rm min}$ is the minimum of the potential. We set
$V(r)=V_{\rm min}$ in the interval $a<r<b$, where $b$ is the
zero-energy turning point of the carbon hard wall. Finally, we let
$V=\infty$ for $r>b$.

For the case $\epsilon < V_0$ considered here, the radial wave
function has the general form
\begin{equation}
f(r)=\left\{ \begin{array}{ll}
 A I_{\nu}(\kappa r) & \mbox{for $r<a$},\\
 B \left[J_{\nu}(\chi r)-N_{\nu}(\chi r) \frac{J_{\nu}(\chi
b)}{N_{\nu}(\chi b)}\right], & \mbox{for $a<r<b$}.
 \end{array}
\right.
\end{equation}
where $J_{\nu}$, $N_{\nu}$ are ordinary Bessel functions, and $I_{\nu}$ are 
modified Bessel functions. Here, $|\epsilon - V_0| = \hbar^2
\kappa^2/2m$, and $\epsilon + V_{\rm min} =\hbar^2 \chi^2/2m$.
Using the matching conditions, the eigenvalue problem reduces to
the solution of the transcendental equation:
\begin{equation}
\frac{\kappa I^{\prime}_{\nu}(\kappa a)}{I_{\nu}(\kappa a)} = 
\frac{\chi \left[J^{\prime}_{\nu}(\chi a) \, N_{\nu}(\chi b) -
N^{\prime}_{\nu}(\chi a) \, J_{\nu}(\chi b)\right]}{J_{\nu}(\chi a)
\, N_{\nu}(\chi b) - N_{\nu}(\chi a)\, J_{\nu}(\chi b)},
\end{equation}
where the prime refers to a derivative with respect to $r$.

\begin{figure}
\caption {Scanning tunneling microscope picture of a single-wall 
nanotube, revealing the hexagonal atomic structure. The tube's radius is 
14 \AA. Adapted from the original (color) version which appeared in Ref. [8].}
\label{fig:nanotubes}
\end{figure}

\begin{figure}
\caption {The model potential (---) and the true potential (--- ---) for
$^3$He inside a single-wall nanotube with R=5 \AA. The ground state
energy ($\cdot \cdot \cdot$) and the first radially excited eigenstate 
(--- $\cdot$ ---), as well as the second excited azimuthal energy level 
(-- --) are also represented (the first excited azimuthal energy level is 
indistinguishable from the ground state on this scale). The wave functions 
(unnormalized) shown correspond to the ground state ($\circ$), the first 
azimuthally excited state ($\Box$) and the first radially excited state 
(+).}
\label{fig:pot}
\end{figure}

\begin{figure}
\caption {The density of states (Eq.~(\ref{eq:dens_states})) for
$^3$He inside a single-wall tube of radius 5 \AA (---) compared
with the 2D density of states
 for $^3$He on a flat surface (--- ---). n$_{2D}$ is the density of
states on a flat surface with the same area $\cal A$, n$_{2D}$= m
$\cal A$/ (2 $\pi 
\hbar^2$). The arrows indicate the three Fermi energies considered.}
\label{fig:dens_states}
\end{figure}

\begin{figure}
\caption {Crossover of specific heat from that of a 1D gas at low
$T$ ($c = 1/2$) to that of a 2D gas at high $T$ ($c = 1$) in the
case of
$^3$He inside a single-wall nanotube of radius 5 \AA\ (---) and 8
\AA\ (--- ---).}
\label{fig:heat_c}
\end{figure}

\begin{figure}
\caption {The 1D and 2D regimes for $^3$He inside a single wall
nanotube. The solid line represents a characteristic azimuthal
excitation temperature estimated by $T \simeq \hbar^2/[2 m k_B
(R')^2]$ (Ref. [4]), where $R'$ is the mean distance of the atom
from the center of the tube. Actual azimuthal excitation energies
($\blacksquare$) and radial excitation energies ($\blacklozenge$,
fitted by dotted line) for radii of 5 \AA\ and 8 \AA\ are shown.
The temperature was expressed in Kelvin. The circle represents the 
experimental [35] vibrational excitation energy for $^3$He adsorbed
on a flat surface of graphite.}
\label{fig:dimens}
\end{figure}

\begin{figure}
\caption {The specific heat as a function of T for noninteracting $^3$He 
(R=5 \AA) with densities N$_1$ (---), N$_2$ (--- ---), and 
N$_3$ (--- $\cdot$ ---), respectively, as defined in text, as well as the 
case of a purely 1D system ($\cdot \cdot \cdot$).}
\label{fig:deg_Fermi}
\end{figure}

\begin{figure}
\caption {The energy per particle of 1D $^4$He as a function of
density at
$T=0$. The inset shows the low density regime. (Reproduced from Ref. [7]).}
\label{fig:bound_state}
\end{figure}

\begin{figure}
\caption {The dimensionless ratio (defined in
Eq.~(\ref{eq:heat_cap})) of the phonon heat capacity to the
temperature (---), as a function of the dimensionless temperature
$t$ (defined in Eq.~)\ref{eq:t})). The dotted curve displays the
constant (linear) dependence on $T$ expected for $C/T$ in a 1D (2D)
system. The actual plot reveals a crossover between 1D and 2D
behavior in the nanotubes. (Reproduced from Ref.~[6]).}
\label{fig:phonons_crossover}
\end{figure}

\begin{figure}
\caption {The relation between crossover temperature and the radius of the
cylindrical shell containing the film. The curves correspond to different
phases of He films and the case of a solid Xe film.The scaling temperature $T_R =\hbar s/ (k_B R)$ is seen to
depend on the specific material parameters. (Reproduced from Ref. [6]).}
\label{fig:phonons_heat_c}
\end{figure}

\newpage
\begin{figure}[ht]
\epsfysize=3.5in \epsfbox{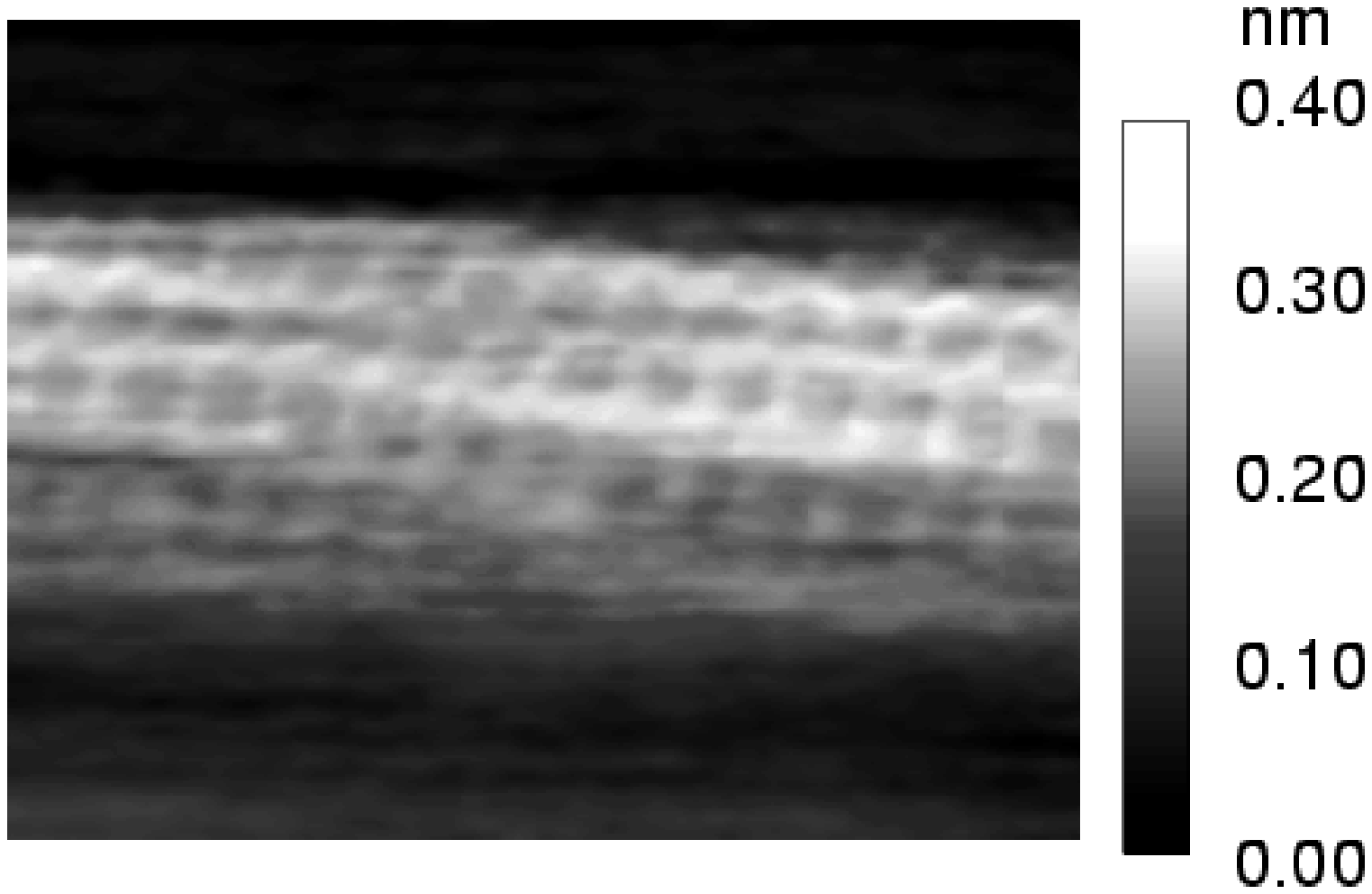}
\end{figure}
\begin{center}
{\bf FIG. 1}
\end{center}

\newpage
\begin{figure}[ht]
\epsfysize=3.5in \epsfbox{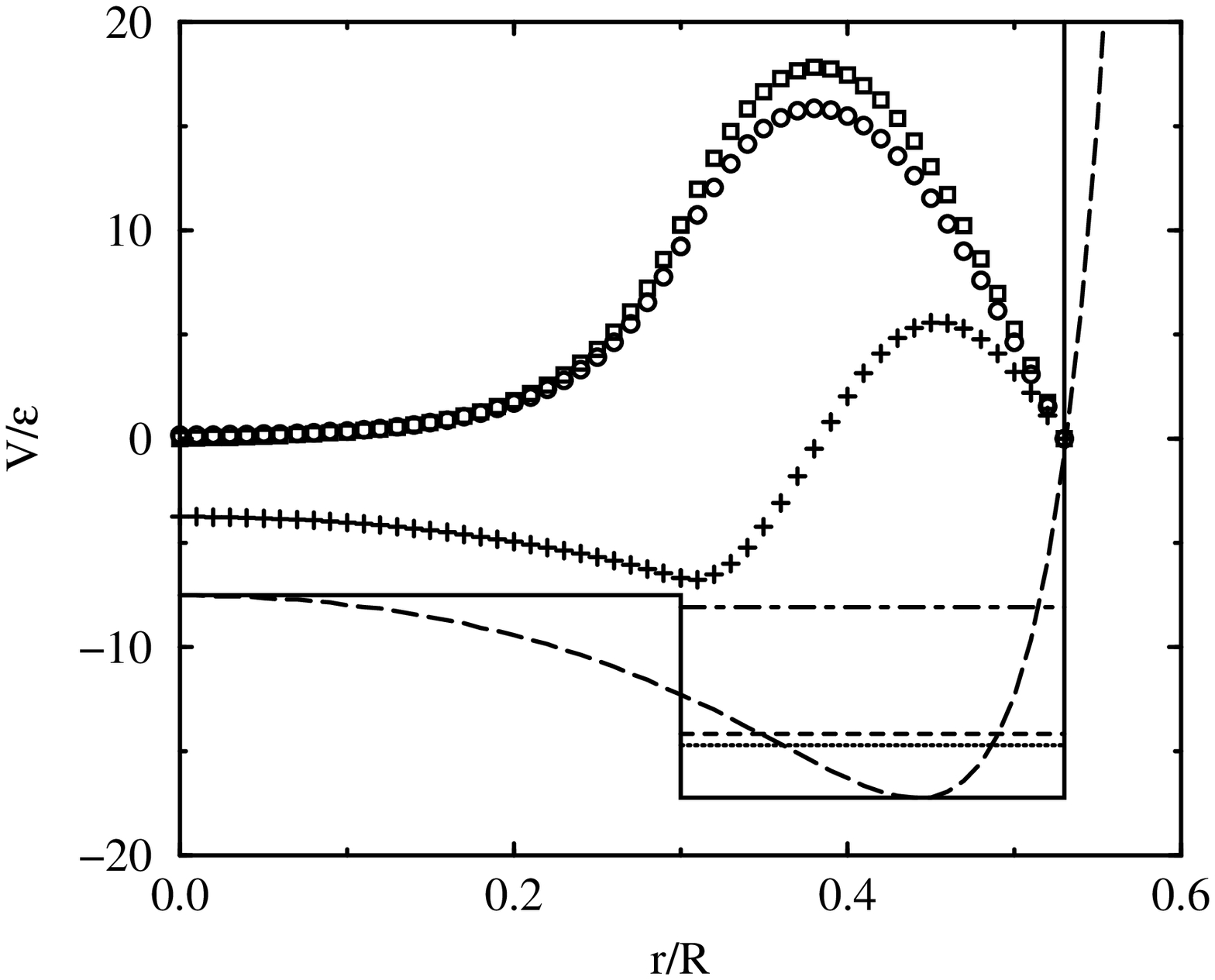}
\end{figure}
\begin{center}
{\bf FIG. 2}
\end{center}

\newpage
\begin{figure}[ht]
\epsfysize=3.5in \epsfbox{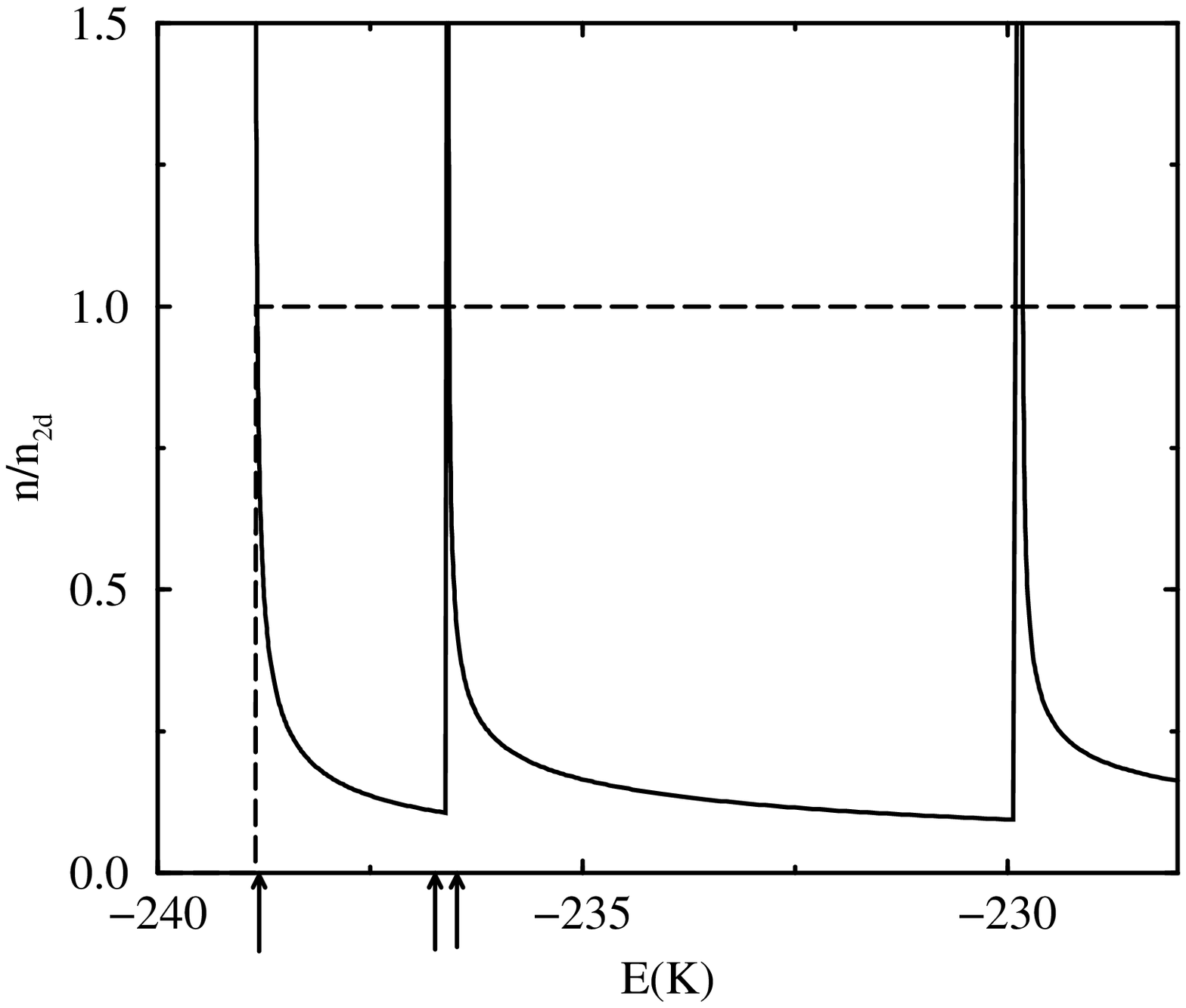}
\end{figure}
\begin{center}
{\bf FIG. 3}
\end{center}

\newpage
\begin{figure}[ht]
\epsfysize=3.5in \epsfbox{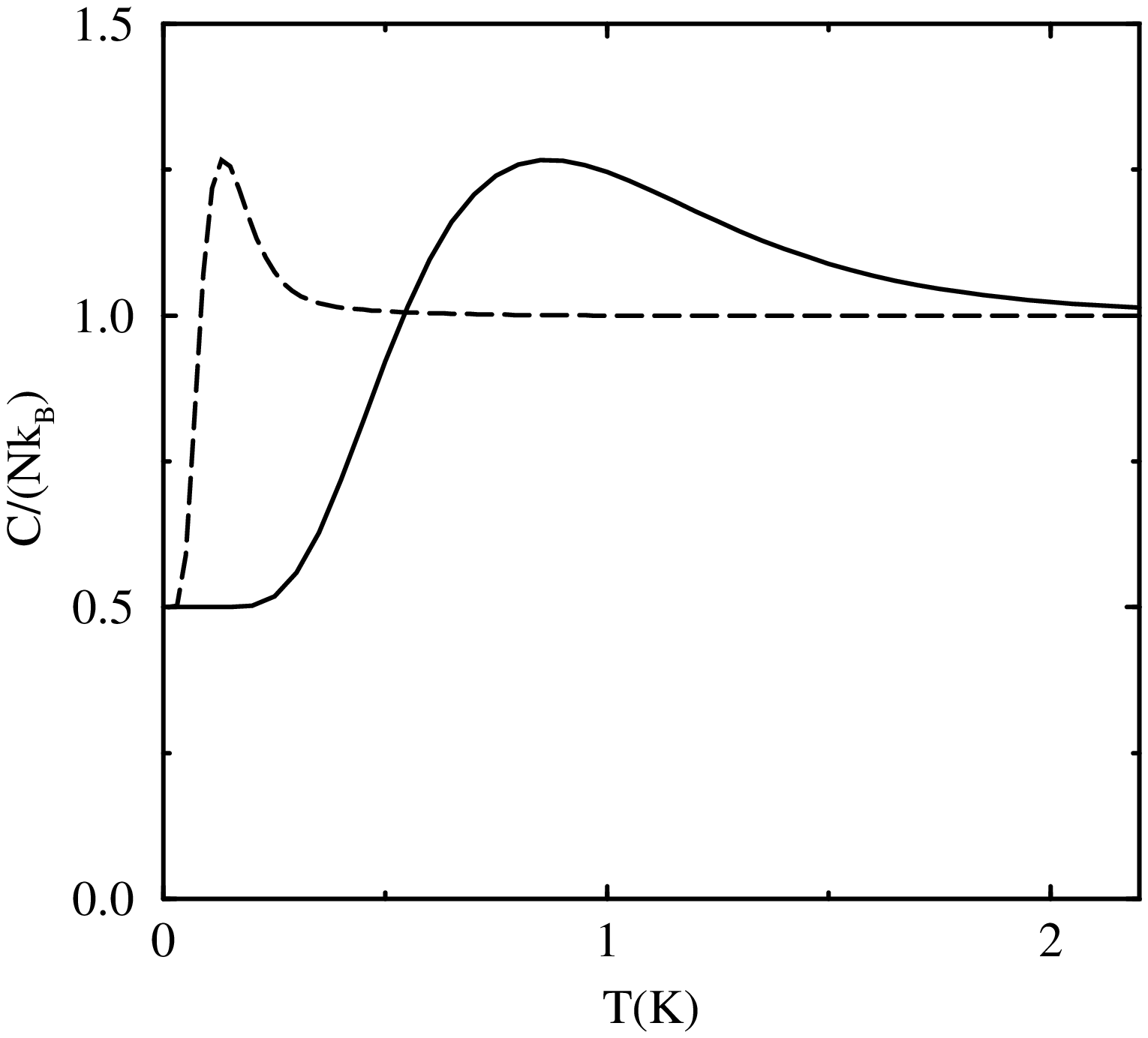}
\end{figure}
\begin{center}
{\bf FIG. 4}
\end{center}

\newpage
\begin{figure}[ht]
\epsfysize=3.5in \epsfbox{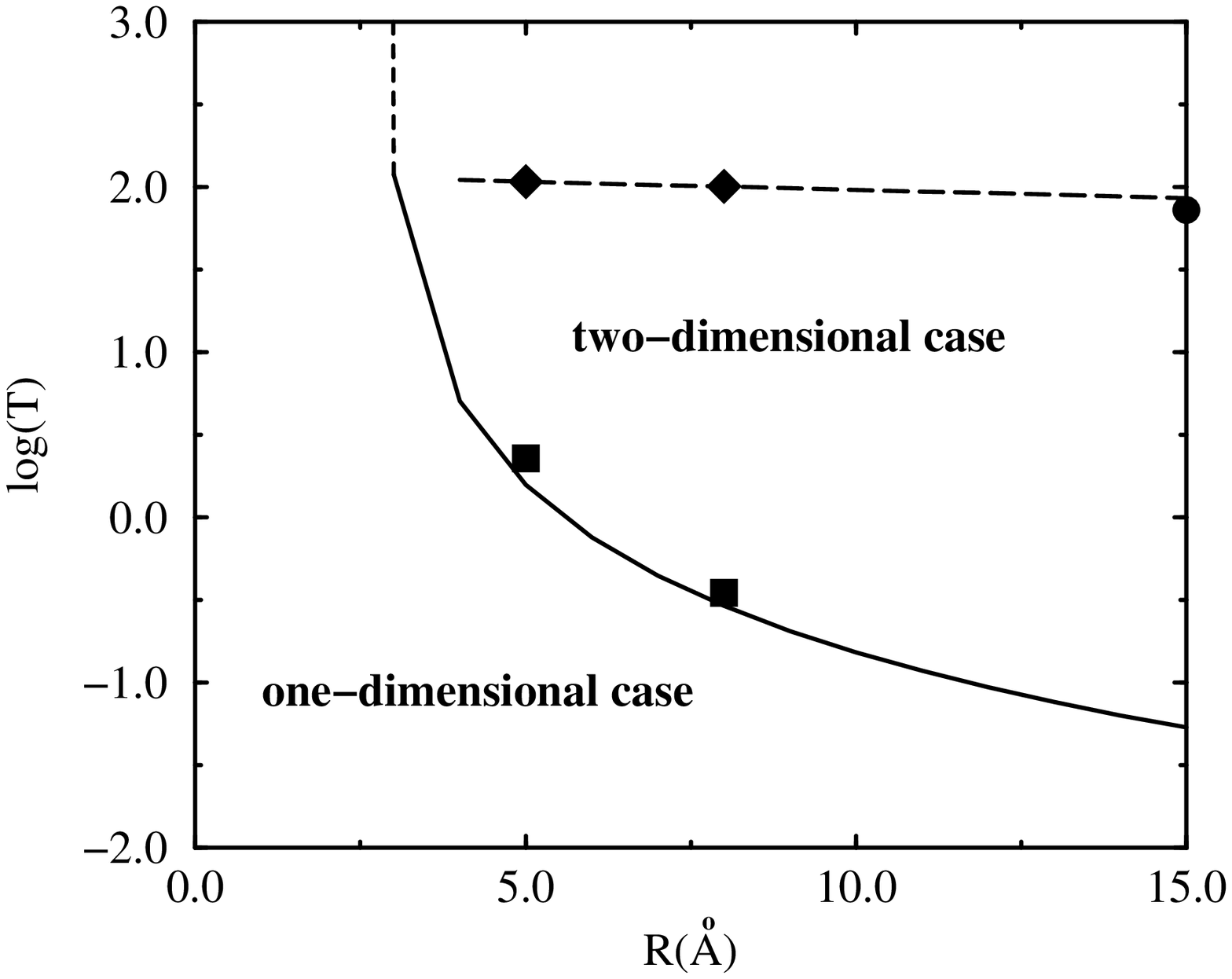}
\end{figure}
\begin{center}
{\bf FIG. 5}
\end{center}

\newpage
\begin{figure}[ht]
\epsfysize=3.5in \epsfbox{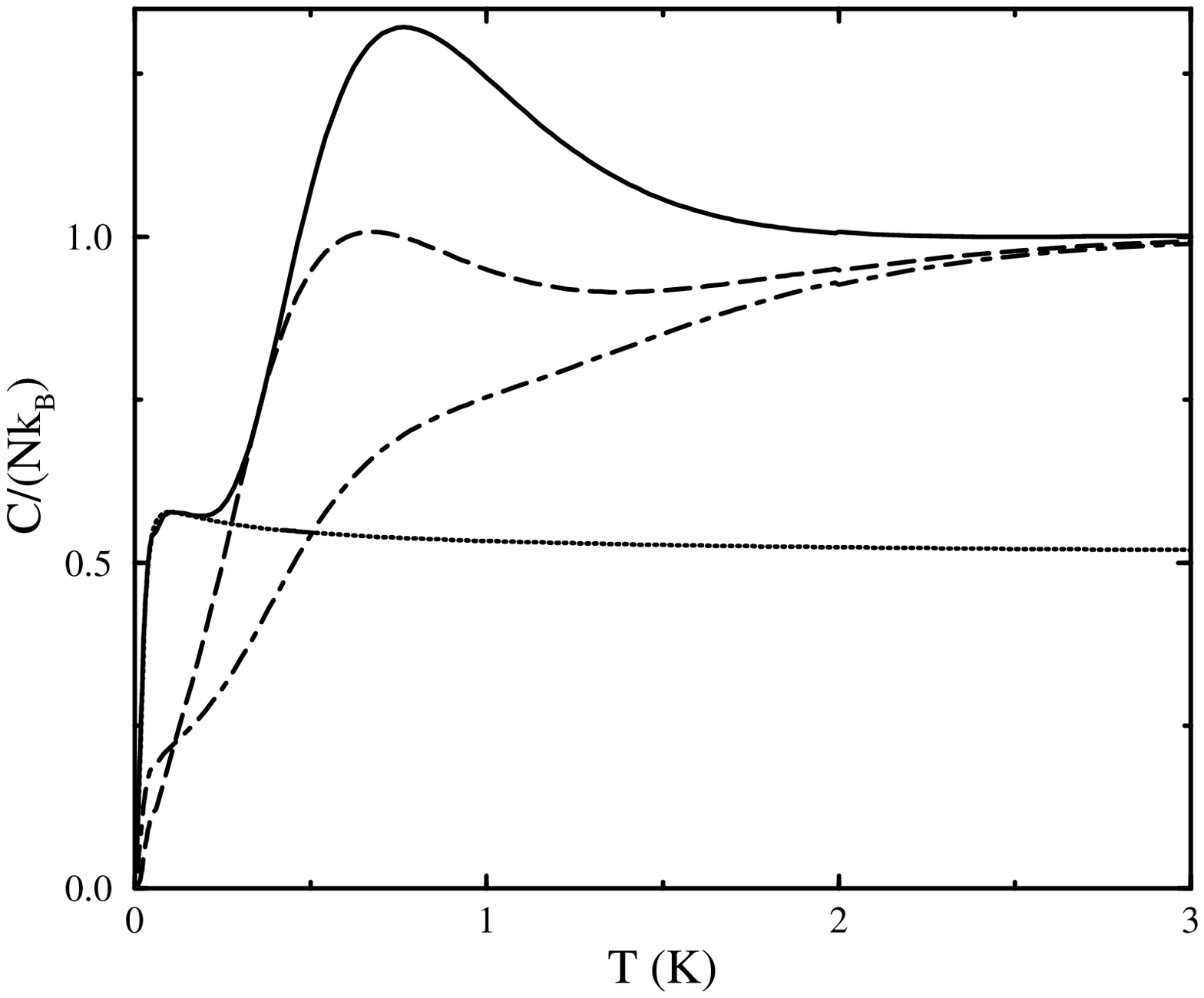}
\end{figure}
\begin{center}
{\bf FIG. 6}
\end{center}

\newpage
\begin{figure}[ht]
\epsfysize=3.5in \epsfbox{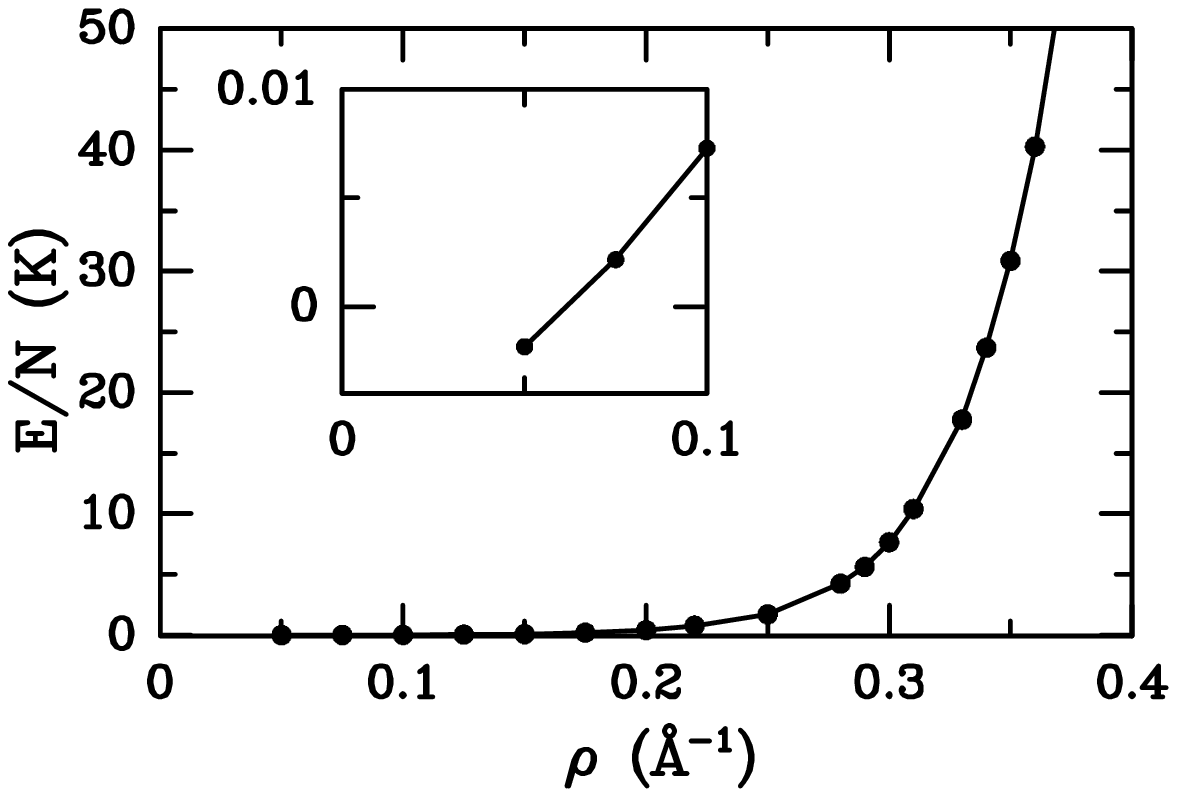}
\end{figure}
\vspace{1in}
\begin{center}
{\bf FIG. 7}
\end{center}

\newpage
\begin{figure}[ht]
\epsfysize=5in \epsfbox{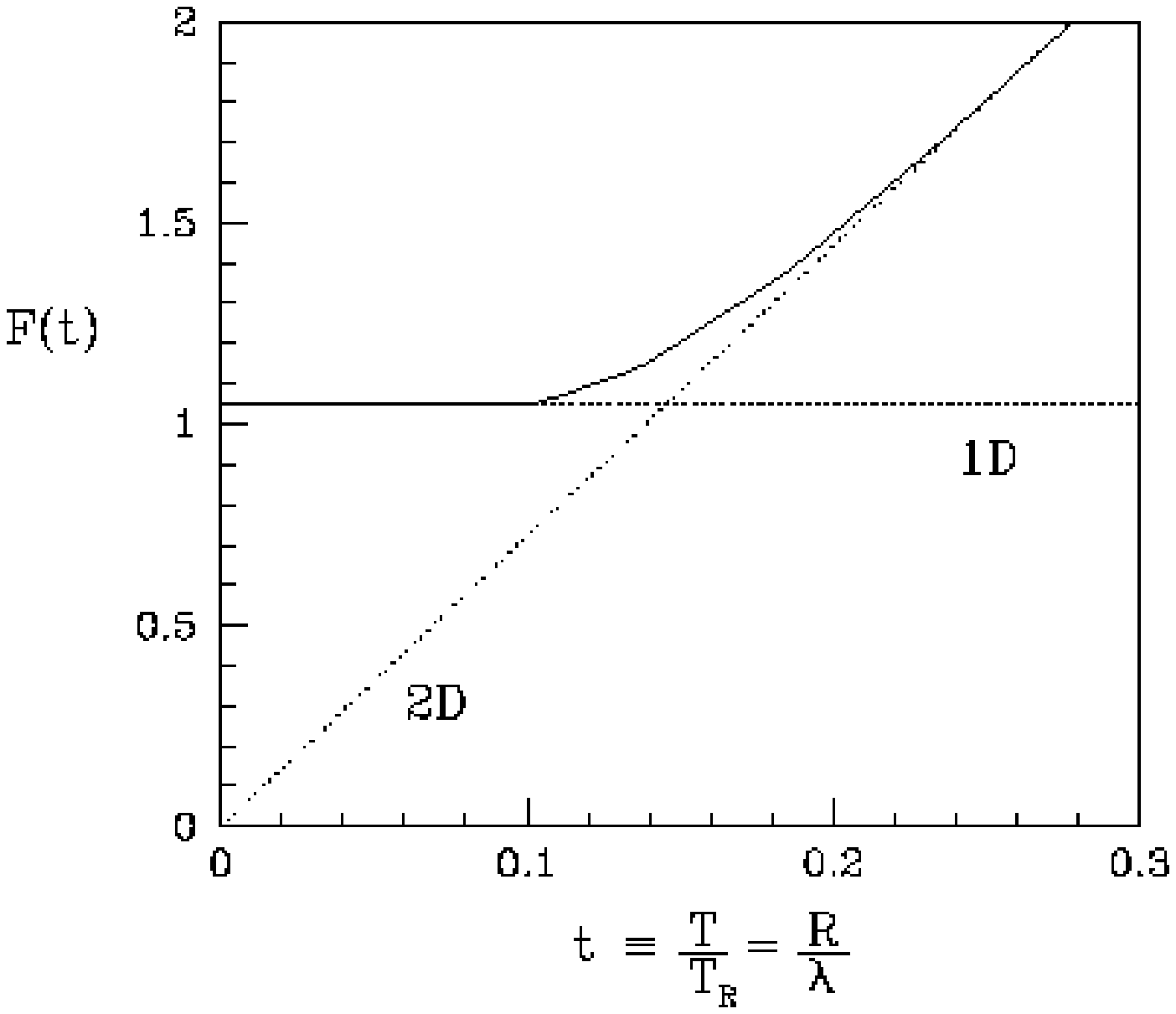}
\end{figure}
\vspace{-3cm}
\begin{center}
{\bf FIG. 8}
\end{center}

\newpage
\begin{figure}[ht]
\epsfysize=5in \epsfbox{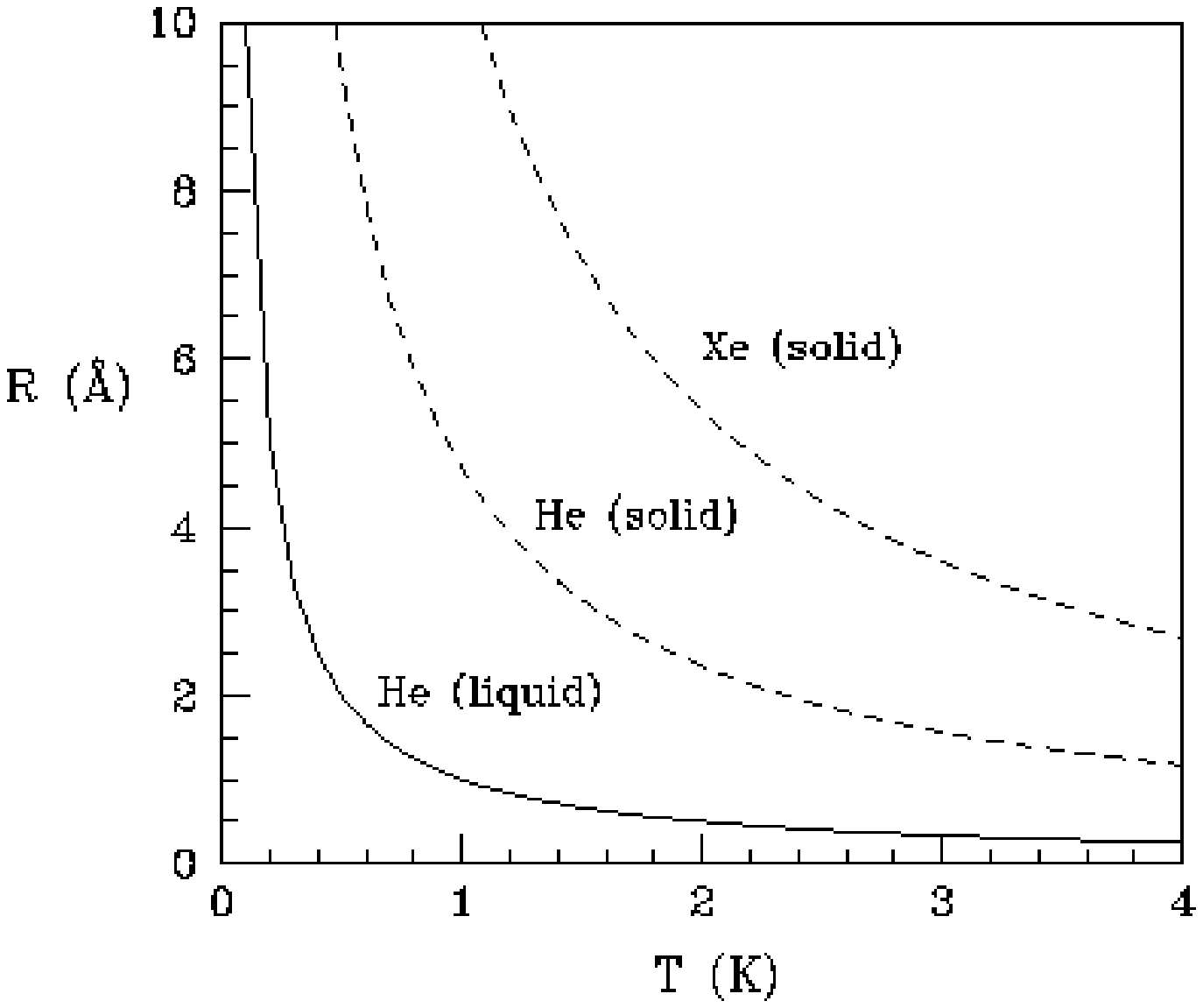}
\end{figure}
\vspace{-3cm}
\begin{center}
{\bf FIG. 9}
\end{center}


\begin{thebibliography}{50}
\bibitem{dos}In 1D, the number of $k$ values in an interval $[k,
k+dk]$ is $(L/2 
\pi) dk$, corresponding to an energy interval $dE= \hbar^2 k dk/m$.
Taking into account degenerate states of negative wave vector $k$,
the density of states (number per unit energy) is $mL/(2 \pi
\hbar^2 k)$, which yields Eq.~(\ref{eq:1d_dos}).
\bibitem{note3}The excitation spectrum for the linear rotator (for
example, the rigid diatomic molecule) has a similar form ($E
\propto l(l+1)$) and exhibits an overshoot of the heat capacity
above $C/Nk_B=1$.
\bibitem{quantum_KE}The quantum kinetic energy per particle, due to a 
particle's confinement by its neighbors at distance $\sim \sigma$, is 
$\sim h/(m \sigma^2)$, while the potential energy is of order $\epsilon$.
\bibitem{note4}Bucatini is a pasta that has a hole along the axis
of each strand.
\end{thebibliography}
\end{document}